\documentclass[conference]{IEEEtran}
\IEEEoverridecommandlockouts
\usepackage{cite}
\usepackage{amsmath,amssymb,amsfonts}
\usepackage{graphicx}
\usepackage{textcomp}
\usepackage{xcolor}
\usepackage{comment}
\usepackage[T1]{fontenc}
\usepackage[utf8]{inputenc}
\usepackage{dblfloatfix} 
\usepackage{algorithm}
\usepackage{algpseudocode}
\usepackage{subcaption}
\usepackage{soul}
\soulregister\mathcal{1}
\soulregister\subseteq{1}
\usepackage{xcolor}
\usepackage{graphicx}
\def\BibTeX{{\rm B\kern-.05em{\sc i\kern-.025em b}\kern-.08em
    T\kern-.1667em\lower.7ex\hbox{E}\kern-.125emX}}
    \usepackage{subcaption}

\begin{document}

\title{Personalized Privacy-Aware MARL Offloading in the Edge-Cloud Continuum }\vspace{-10mm}
\author{
\IEEEauthorblockA{
Muhammad Rafid$^{\dagger,\mathsection}$,
Golshan Famitafreshi$^{\ddagger}$,
Kostas Ramantas$^{\ddagger}$,
Luis Alonso$^{\mathsection}$,
Christos Verikoukis$^{\dagger,*}$
}
\IEEEauthorblockA{
$^{\dagger}$ISI/ATHENA RC, 26504 Patra, Greece 
(e-mail: \{mrafid, cveri\}@isi.gr)\\
$^{\mathsection}$Universitat Polit\`ecnica de Catalunya (UPC), Barcelona, Spain (e-mail: \{luis.g.alonso\}@upc.edu)\\
$^{*}$University of Patras, 26504 Patra, Greece\\
$^{\ddagger}$Iquadrat Informatica S. L., Barcelona, Spain 
(e-mail: \{g.famitafreshi, kramantas\}@iquadrat.com)
}
}
\maketitle
\begin{abstract}
Task offloading in End--Edge--Cloud computing enables resource-constrained User Devices (UDs) to execute computation-intensive applications with reduced latency and energy consumption. However, existing task offloading schemes generally overlook the heterogeneous privacy requirements of users, leading to inefficient privacy-aware offloading decisions. In this paper, we propose a personalized privacy-aware task offloading framework, where the privacy requirement of each task is quantified through a User-Centric Privacy (UCP) model that considers task sensitivity, user privacy preferences, and task urgency. The task offloading problem is formulated as a Markov Decision Process (MDP) to jointly optimize the offloading ratio and the target computing server while minimizing the long-term system cost under latency, energy, and privacy constraints. To efficiently solve the resulting high-dimensional continuous optimization problem, we develop an Advantage-Weighted Actor-Critic (AWAC)-enhanced Proximal Policy Optimization (PPO)-based Multi-Agent Reinforcement Learning (MARL) algorithm under the Centralized Training and Decentralized Execution (CTDE) paradigm. Simulation results demonstrate that the proposed framework achieves lower average system cost, lower energy consumption, and higher task success rate than representative baselines under diverse privacy requirements.
\end{abstract}
\section{Introduction}
In recent years, the rapid growth of intelligent mobile devices and latency-sensitive applications, such as online gaming, virtual reality, and autonomous driving, has significantly increased the demand for computation-intensive services~\cite{khan2019edge}. To address this, Multi-Access Edge Computing (MEC) deploys Edge Servers (ESs) close to users, enabling resource-constrained devices to offload computational workloads and thereby reducing task execution latency~\cite{dong2024task}. However, the offloading process often requires transmitting user-related information, such as task characteristics, communication patterns, and location-dependent data, which may expose sensitive information and create privacy leakage risks~\cite{min2025personalized}. Consequently, achieving efficient task offloading while preserving user privacy has become a critical challenge in dynamic MEC-enabled networks \cite{alzailaa2026resource}.

Most existing privacy-preserving task offloading studies protect user information through privacy-aware task allocation, location privacy protection, and personalized privacy-preserving mechanisms~\cite{ni2019enabling}. Although these approaches effectively reduce privacy leakage, they generally adopt a uniform protection strategy and overlook the heterogeneous privacy requirements of individual users~\cite{liu2020comp}. In practice, however, different users require different levels of privacy protection depending on their preferences and application demands. Thus, applying identical mechanisms to all users may either expose sensitive information or cause unnecessary performance degradation. To address this challenge, recent studies have emphasized personalized privacy protection, in which privacy requirements are dynamically adjusted according to user-specific characteristics~\cite{li2024personalized}. Therefore, incorporating personalized privacy into offloading decisions has become a promising approach for balancing privacy protection and system performance in edge--cloud systems.

A large variety of studies have developed privacy-aware task offloading and resource management strategies in MEC systems. However, most rely on centralized decision-making, where privacy requirements, resource availability, and offloading decisions are processed by a central controller~\cite{liu2020comp,li2024personalized}, which introduces excessive communication overhead and becomes impractical in large-scale and dynamic MEC environments. To overcome this, recent studies have shown increasing interest in decentralized task offloading frameworks~\cite{lin2024decentralized}, mainly implemented through MARL approaches such as MADDPG-based~\cite{guo2022multi}, and COMA-based~\cite{liu2020distributed} methods. Most such studies adopt an actor--critic architecture with CTDE, where a centralized critic utilizes global information during training while each actor executes decisions using only local observations \cite{liu2020distributed}, enabling scalable and adaptive decision-making under dynamic network conditions.

\begin{figure}[t]
\begin{center}
{{\includegraphics[width=\linewidth, height=6.4cm,]{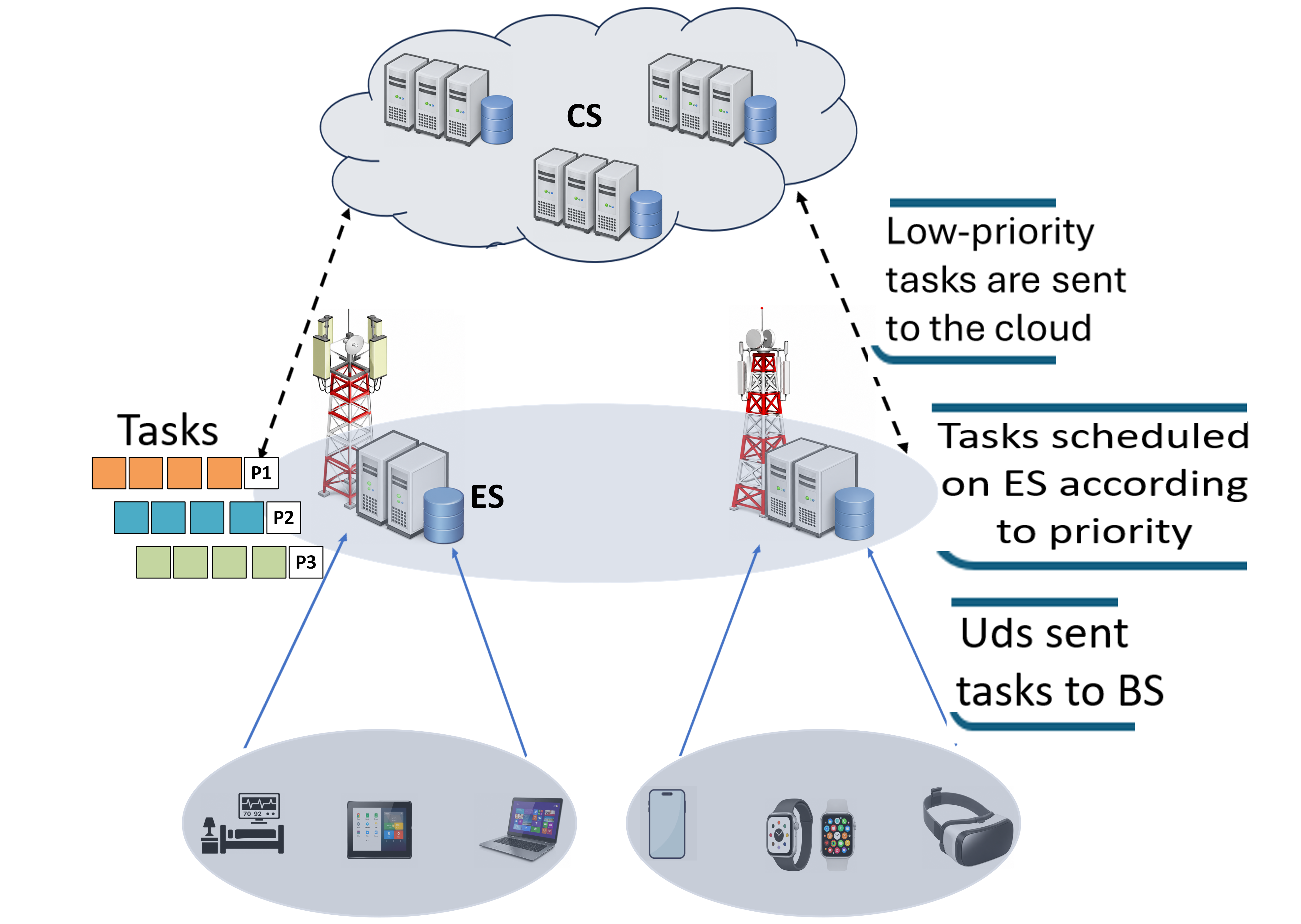 }}}
  \caption{System model of the proposed personalized privacy-aware task offloading framework in an End--Edge--Cloud continuum, where UDs perform partial task offloading to ESs, while low-priority tasks are forwarded to the CS for execution.} \vspace{-3mm}
  \label{spectrum diagram}
  \end{center}
\end{figure}
Motivated by the above observations, we propose a personalized privacy-aware task offloading framework for End--Edge--Cloud computing environments. The proposed framework explicitly incorporates User-Centric Privacy (UCP) requirements into the offloading decision process through task sensitivity, user privacy preference, and task urgency~\cite{li2024personalized}. To address the resulting continuous optimization problem, we develop an AWAC-enhanced PPO-based MARL framework operating under the CTDE paradigm~\cite{chandler2025network}. Unlike DQN-based methods designed for discrete action spaces~\cite{li2018task}, the proposed method supports continuous offloading decisions, enabling each user to autonomously determine its offloading strategy from local observations under dynamic network conditions.

The main contributions of this paper can be summarized as follows.
\begin{itemize}
    \item We propose a personalized, privacy-aware computation offloading framework for End--Edge--Cloud computing, in which the privacy requirement of each task is dynamically quantified using a UCP model, and formulate an optimization problem that jointly minimizes system latency and energy consumption while satisfying heterogeneous user privacy requirements.
    \item We formulate the task offloading problem as a continuous MDP and develop an AWAC-enhanced PPO-based MARL framework under the CTDE paradigm, enabling each user to autonomously determine the optimal task partitioning ratio and target computing server from local observations while ensuring stable policy learning.
    \item Extensive simulations demonstrate that the proposed framework consistently outperforms representative MARL-based baselines in terms of average system cost, energy consumption, task success rate, and convergence stability under diverse network and privacy conditions.
\end{itemize}
The remainder of this paper is as follows: Section~\ref{sec:system model} presents the system model, while the problem formulation and proposed MARL framework is presented in Section~\ref{sec:problem formulation}. The simulation results are presented in Section~\ref{sec:results}, followed by concluding remarks and future research directions (Section~\ref{sec:conclusion}).
\section{System Model}
\label{sec:system model}
 The personalized privacy-aware task offloading model proposed in this paper is constructed in an End–-Edge–-Cloud environment, where ESs and cloud computing servers provide computing services to a large number of UDs generating computation tasks. Each user can offload complex tasks to a nearby ES or cloud server (CS) according to specific offloading strategies, thereby reducing energy consumption and task execution delay. The proposed framework incorporates user-specific privacy requirements into the offloading decision process for multi-user, multi-task computing scenarios.

As shown in Fig.~\ref{spectrum diagram}, we consider a personalized privacy-aware End--Edge--Cloud architecture consisting of three layers: the user layer, the edge layer, and the cloud layer. Let $\mathcal{S}=\{1,\ldots,S\}$ and $\mathcal{U}=\{1,\ldots,U\}$ denote the sets of ESs and UDs, respectively, and let $\mathcal{N}=\{1,\ldots,N\}$ be the set of decision slots. UDs are randomly distributed across the coverage areas of ESs, and the system operates in a slot-based manner. Each ES $s\in\mathcal{S}$ is characterized by a computing capacity $f_s$ and maintains a task queue $q_s$, where $q_s^n$ denotes the queued workload at slot $n$. The CS provides substantial computational resources and is characterized by a computing capacity $f_c$, where $f_c \gg f_s$. Each UD $(u\in\mathcal{U})$ is equipped with a local processor of capacity $f_u$ and an energy budget $E_u^{\max}$, with $E_u^n$ denoting the available energy at slot $n$.

At every slot $n$, UD $u$ generates a computation-intensive and latency-sensitive task
$J_u^n=(C,L_u^n,D^{\max})$,
where $C$ denotes the required CPU cycles, $L_u^n$ represents the input data size, and $D^{\max}$ is the maximum tolerable delay. Each task is further associated with a sensitivity level, a user privacy preference, and an urgency requirement, which collectively influence the offloading decision process. Each UD maintains an offloading controller $\pi_u$ that determines, at every slot, how the task $J_u^n$ is partitioned among the device, a selected ES, and the CS. We define the offloading ratio decision set as
\begin{equation}
x_u^n={x_{u,u}^n,x_{u,s}^n}, \qquad \forall u\in\mathcal{U}, s\in\mathcal{S},
\end{equation}
where $x_{u,u}^n\in[0,1]$ indicates the proportion of $J_u^n$ processed locally on UD $u$, $(1-x_{u,u}^n)x_{u,s}^n\in[0,1]$ represents the proportion offloaded to the selected ES $s$, and $(1-x_{u,u}^n)(1-x_{u,s}^n)\in[0,1]$ denotes the proportion offloaded to the CS. This partial offloading enables flexible task allocation and parallel execution across the user, edge, and cloud layers, improving system responsiveness and resource utilization. This offloading model applies to a wide range of latency-sensitive applications, such as edge-based malware detection, where execution traces are partially offloaded to secure ES or CS for real-time scanning against signature files, thereby reducing the computational burden on UDs \cite{burguera2011crowdroid}. Due to the absence of direct Device-to-Device (D2D) communications in most practical scenarios, the offloading controllers operate independently but collectively aim to minimize the system cost.
\subsection{Communication Model}
The wireless communication link between UD $u$ and ES $s$ is modeled using a block fading channel~\cite{meng2019power}. Accordingly, the uplink channel gain over the offloading link $L_{u,s}^{n}$ at slot $n$ is expressed as $h_{u,s}^{n}=\alpha_{u,s}^{n}\left|g_{u,s}^{n}\right|^{2}$, where $\alpha_{u,s}^{n}$ accounts for the large-scale fading effects, including path loss and shadowing, while $g_{u,s}^{n}$ denotes the small-scale fading component. The large-scale fading coefficient is given by
\begin{equation}
\alpha_{u,s}^{n}
=
-148.1
-
37.6\log_{10}(d_{u,s}^{n})
+
10\log_{10}(z),
\end{equation}
where $d_{u,s}^{n}$ represents the distance between UD $u$ and ES $s$, and $z$ is a log-normal random variable accounting for shadowing effects. The small-scale fading component follows the Jakes fading model \cite{dent1993jakes}.

Assuming that multiple UDs simultaneously offload tasks within the same time slot, the aggregate uplink interference observed at ES $s$ can be expressed as
\begin{equation}
I_{s}^{n}
=
\sum_{u' \in \mathcal{U},\,u' \neq u}
\mathbf{1}(x_{u',s}^{n}>0)
p^{tx}
h_{u',s}^{n},
\end{equation}
where $x_{u',s}^{n}$ denotes the edge-offloading ratio of UD $u'$, $p^{tx}$ denotes the transmission power of the UD, and $\mathbf{1}(\cdot)$ represents the indicator function. Accordingly, the achievable transmission rate between UD $u$ and ES $s$ at slot $n$ is given by
\begin{equation}
R_{u,s}^{n}
=
B
\log_{2}
\left(
1+
\frac
{p^{tx}h_{u,s}^{n}}
{I_{s}^{n}+\sigma^{2}}
\right),
\end{equation}
where $B$ denotes the available channel bandwidth and $\sigma^{2}$ represents the additive white Gaussian noise power.

In addition to the wireless end user-to-edge communication, tasks assigned to the CS must be transmitted through the backhaul link connecting ESs and the cloud. Since this connection is typically supported by high-speed and stable network infrastructure, the transmission rate between ES $s$ and the CS is assumed to be constant and is denoted by $R_{s,c}$. Therefore, the edge-to-cloud transmission delay for the task portion forwarded to the cloud is given by
\begin{equation}
t_{u,s}^{s2c,n}
=
\frac{(1-x_{u,u}^{n})(1-x_{u,s}^{n})L_{u}^{n}}
{R_{s,c}},
\end{equation}
where $R_{s,c}$ denotes the transmission rate of the backhaul link between ES $s$ and the CS, and $(1-x_{u,u}^{n})(1-x_{u,s}^{n})L_{u}^{n}$ represents the amount of task data forwarded from the ES to the CS.

\subsection{Personalized User-Centric Privacy Model}

Different users exhibit varying privacy requirements across computation offloading scenarios. A uniform privacy protection strategy may not adequately satisfy users with strict privacy concerns, while users with lower privacy requirements may prioritize service performance and task execution efficiency. Therefore, a personalized UCP model is developed to capture the heterogeneous privacy requirements of individual users and incorporate them into the offloading decision process. Specifically, the privacy requirement of a user is characterized by three key attributes, namely task sensitivity, privacy preference, and task urgency, which collectively influence task priority within the proposed End--Edge--Cloud framework.

\emph{1) Task Sensitivity:} Offloading tasks involving highly sensitive information, such as healthcare records, financial transactions, and personal user data, requires stronger privacy protection to prevent the leakage of critical information. Therefore, tasks with higher sensitivity levels demand stricter privacy preservation during the offloading process. Following the UCP paradigm, the task sensitivity score $S_u^n$ is determined based on the access duration $M$ and access frequency $F$ of sensitive data, and is expressed as
\begin{equation}
S_u^n =\alpha_M M+\alpha_F F,
\end{equation}
where $\alpha_M$ and $\alpha_F$ denote the weighting factors associated with access duration and access frequency, respectively, satisfying
$\alpha_M+\alpha_F=1$.

\emph{2) Privacy Preference:} Privacy preference reflects the user's desired level of privacy protection during task offloading, where users with stronger preferences prioritize privacy preservation and others focus more on service performance. The privacy preference score is defined as\vspace{-2mm}
\begin{equation}
P_u^n = 2\theta_u^n - 1,
\end{equation}
where $\theta_u^n \in [0,1]$ denotes the privacy preference indicator. A larger value of $P_u^n$ corresponds to a stronger privacy requirement.

\emph{3) Task Urgency:} Task urgency reflects the time sensitivity of a computation task. Tasks with stringent latency requirements, such as real-time monitoring, autonomous control, and emergency response, require rapid execution, whereas delay-tolerant tasks can accommodate longer execution times and higher privacy overhead. The task urgency is denoted by $T_u^n$, where a larger value indicates a higher urgency level and greater emphasis on timely execution.

The personalized UCP score of UD $u$ at slot $n$ is defined as\vspace{-2mm}
\begin{equation}
UCP_u^n = w_1 S_u^n + w_2 P_u^n - w_3 T_u^n,
\end{equation}
where $w_1$, $w_2$, and $w_3$ denote the weighting factors associated with task sensitivity, privacy preference, and task urgency, respectively, satisfying
$w_1+w_2+w_3=1$.

A higher $UCP_u^n$ indicates a stronger privacy requirement, whereas a lower value implies the task can tolerate weaker privacy protection and prioritize computational efficiency.
\subsection{Local Computing Model}
The partial offloading strategy is applied to UD  $u$'s task $J_u^n$ at slot $n$. If $x_{u,u}^{n}>0$, the corresponding fraction of the task is executed locally. Accordingly, the local computation delay and energy consumption are given by
\begin{equation}
t_{u}^{\mathrm{loc},n}
=
\frac{x_{u,u}^{n}\,C\,L_{u}^{n}}
{f_{u}},
\qquad
e_{u}^{\mathrm{loc},n}
=
\kappa_{u}\,f_{u}^{2}\,x_{u,u}^{n}\,C\,L_{u}^{n},
\label{eq:local-lat-energy}
\end{equation}
where $f_u$ is the CPU frequency of UD $u$ and $\kappa_u$ is the energy coefficient of the device processor.
\vspace{-2mm}
\subsection{Offloaded Computing Model}

\vspace{-1mm}\emph{1) Edge Offloading:} If $(1-x_{u,u}^{n})x_{u,s}^{n}\neq0$, the fraction $(1-x_{u,u}^{n})x_{u,s}^{n}$ of task $J_u^{n}$ is offloaded to the selected ES $s$ for remote execution in slot $n$. The end-to-end latency comprises three components: (i) the transmission time required to deliver $(1-x_{u,u}^{n})x_{u,s}^{n}L_u^{n}$ bits over the link $\mathcal{L}_{u\rightarrow s}^{n}$, given by $t_{u,s}^{\mathrm{tx},n}=\frac{(1-x_{u,u}^{n})x_{u,s}^{n}L_u^{n}}{R_{u,s}^{n}}$; (ii) the waiting time at ES $s$ due to the queued workload and previously arrived offloaded tasks, denoted by $t_{u,s}^{\mathrm{wait},n}$; and (iii) the execution time of the offloaded task portion at the ES CPU, given by $t_{u,s}^{\mathrm{exe},n}=\frac{(1-x_{u,u}^{n})x_{u,s}^{n}L_u^{n}C}{f_s}$. Accordingly, the edge-side delay and offloading energy consumption are
\begin{equation}
t_{u,s}^{\mathrm{edge},n}
=
t_{u,s}^{\mathrm{tx},n}
+
t_{u,s}^{\mathrm{wait},n}
+
t_{u,s}^{\mathrm{exe},n},
\end{equation}
\begin{equation}
e_{u,s}^{\mathrm{edge},n}
=
p^{tx}\,
t_{u,s}^{\mathrm{tx},n}.
\end{equation}
\emph{2) Cloud Computing Model:} When a fraction of the task is further offloaded to the CS, the total delay includes three components: transmission from the UD to the ES, transmission from the ES to the CS, and cloud-side computation. The total delay for cloud offloading is given as follows:
\begin{equation}
t_{u,c}^{\mathrm{cloud},n}
=
t_{u,s}^{\mathrm{tx},n}
+
(1-x_{u,u}^{n})(1-x_{u,s}^{n})
\left(
\frac{L_u^{n}}{R_{s,c}}
+
\frac{CL_u^{n}}{f_c}
\right),
\end{equation}
where $R_{s,c}$ is the transmission rate between the ES and the CS, and $f_c$ denotes the computing capability of the CS. \vspace{-2mm}
\subsection{System Cost Model}
The considered system cost consists of two components, namely task completion delay and energy consumption.

\emph{1) Task Completion Delay:}
Since task $J_u^n$ can be executed simultaneously across the UD, ES, and CS, the overall task completion delay is determined by the maximum execution time among the three computation modes, which is given by
\begin{equation}
t_u^n
=
\max
\left(
t_u^{\mathrm{loc},n},
t_{u,s}^{\mathrm{edge},n},
t_{u,c}^{\mathrm{cloud},n}
\right).
\end{equation}
The task completion delay should satisfy the latency constraint
\begin{equation}
t_u^n
\le
D^{\max}.
\label{maxtime}
\end{equation}

\emph{2) Energy Cost:} Different from the time cost, the total energy consumption of task $J_u^{n}$ includes local computation energy and wireless transmission energy incurred during task offloading, which is given by
\begin{equation}
e_u^{n}
=
e_u^{\mathrm{loc},n}
+
e_{u,s}^{\mathrm{edge},n}.
\label{eq:energy}
\end{equation}
For each slot $n$, the remaining energy budget of UD $u$ is
\begin{equation}
E_u^{n}
=
E_u^{\max}
-
e_u^{n},
\qquad
E_u^{n}\ge0.
\label{eq:budget}
\end{equation}

\emph{3) System Cost:} The task cost $c_u^{n}$ for executing $J_u^{n}$ is given by $c_u^{n}=\eta\,t_u^{n}+(1-\eta)\,e_u^{n}$, where $\eta$ is the weight of time cost. Hence, the average cost per slot $n$ is calculated as $\Psi^{n}=\frac{1}{U}\sum_{u=1}^{U}c_u^{n}$, and the average system cost $\Psi$ over all $N$ slots is given by
\begin{equation}
\Psi=\frac{1}{N}\sum_{n=1}^{N}\Psi^{n}.
\label{cost}
\end{equation}
\section{Problem Formulation and Proposed MARL Framework} \vspace{-2mm}
\label{sec:problem formulation}
We formulate the personalized privacy-aware task offloading problem as a long-term cost minimization problem and cast it as an MDP solved via the proposed AWAC-enhanced PPO-based MARL framework under CTDE.
\subsection{Optimization Problem Formulation} 
Based on the model introduced above, we formulate the optimization problem for the proposed personalized privacy-aware End--Edge--Cloud continuum framework. Since unreasonable offloading strategies may increase task delay, energy consumption, and privacy risk, we explicitly optimize the local execution ratio and edge offloading ratio. Viewing the decision variables $x_{u,u}^{n}$ and $x_{u,s}^{n}$ as optimization variables, the system cost can be considered as a function of these decisions, i.e., $c_u^{n}(x_{u,u}^{n},x_{u,s}^{n})$. Our objective is to minimize the average system cost over the entire episode.\vspace{-3mm}
\begin{equation}
\mathcal{P}:
\quad
\min_{x_{u,u}^{n},\,x_{u,s}^{n}}
c_u^{n}\!\left(x_{u,u}^{n},x_{u,s}^{n}\right),
\label{pro}
\end{equation}
\begin{equation*}
\text{s.t. Eqs.~}\eqref{maxtime}, \eqref{eq:budget}, \eqref{cost}
\text{ and }
0 \le x_{u,u}^{n},x_{u,s}^{n}\le 1.
\end{equation*}
The above problem is challenging to solve since it has an \textit{NP-hard} nature. The main idea is to draw an analogy between the considered problem and the Generalized Assignment Problem (Section 15 in~\cite{hooker2006operations}). In essence, it reduces to allocating computation tasks among local devices, ESs, and cloud resources such that the available communication and computation capacities are efficiently utilized while the overall system cost is minimized.

According to \eqref{cost}, minimizing $c_u^{n}$ in \eqref{pro} is equivalent to minimizing the accumulated per-slot costs across $N$ slots. Because the system cost is inversely related to the reward, problem $\mathcal{P}$ can be modeled as a discrete-time MDP, which aims to maximize the cumulative reward while minimizing long-term delay and energy consumption under privacy constraints.
\subsection{Proposed AWAC-Enhanced PPO-Based MARL Framework}
Concerning the substantial overhead associated with collecting all UDs' states at every slot and the poor scalability of centralized scheduling, we adopt a decentralized MARL formulation. To enable distributed decision-making under dynamic and partially observable End--Edge--Cloud environments, the proposed privacy-aware task offloading framework is modeled as an MDP, where each UD operates as an autonomous learning agent interacting with its environment. Traditional DQN-based approaches are limited by their discrete action representation and are not suitable for continuous control problems such as task partitioning and resource allocation \cite{li2018task}. Accordingly, we adopt an AWAC-enhanced PPO-based MARL framework to learn adaptive offloading policies that minimize the long-term system cost \cite{chandler2025network}. Following the standard definition of an MDP \cite{shakarami2020survey} $(\mathcal{S},\mathcal{A},\mathcal{T},\mathcal{R})$, each agent observes a local state $s_u^n=\{L_u^n,h_{u,s}^n,q_s^n,E_u^n,UCP_u^n\}$, where $L_u^n$, $h_{u,s}^n$, $q_s^n$, $E_u^n$, and $UCP_u^n$ denote the task size, channel gain, queue length of the selected ES, remaining energy budget, and personalized privacy requirement, respectively. Based on the observed state, the agent determines an action $a_u^n=\{x_{u,u}^n,x_{u,s}^n\}$, where $x_{u,u}^n$ and $x_{u,s}^n$ represent the local execution ratio and edge offloading ratio, respectively. The state transition probability $\mathcal{T}(s^{n+1}\mid s^n,a^n)$ captures the evolution of channel conditions, task arrivals, queue dynamics, and energy availability over time.
The reward plays a crucial role in stabilizing the learning process and guiding agents toward efficient privacy-aware offloading decisions. The episodic reward of each agent is defined as
\begin{equation}
\small
r_u=
\begin{cases}
\dfrac{1}{\bar c_u},
&
\text{if }(V_t=0,\;V_e=0,\;V_p=0),
\\[2pt]
-\left(
\beta_tV_t+
\beta_eV_e+
\beta_pV_p
\right),
&
\text{otherwise},
\end{cases}
\end{equation}
where $\bar c_u$ denotes the average episodic cost of UD $u$, while $V_t$, $V_e$, and $V_p$ represent cumulative delay, energy, and privacy violations, respectively. The coefficients $\beta_t$, $\beta_e$, and $\beta_p$ determine the penalty weights associated with the corresponding violations. A positive reward is granted only when all latency, energy, and privacy requirements are satisfied throughout the episode, whereas any violation yields a negative penalty proportional to the degree of constraint violation.

Under the CTDE paradigm, a centralized critic estimates the state-value function $V(s_u^n)$, while each actor executes decisions using only local observations. The clipped advantage is defined as
\begin{equation}
A_u^n = \mathrm{clip}(R_u^n - V(s_u^n), -c, c),
\label{eq:adv}
\end{equation}
where $R_u^n$ denotes the empirical return of agent $u$ at slot $n$, and $c$ is a clipping threshold introduced for numerical stability. To prioritize high-quality actions, an exponential advantage weight is introduced as
\begin{equation}
w_u^n = \exp(A_u^n/\tau), \qquad
w_u^n \leftarrow \min(w_u^n, w_{\max}),
\label{eq:weight}
\end{equation}
where $\tau$ controls the weighting sharpness and $w_{\max}$ limits excessively large updates. The actor is then optimized using the weighted PPO clipped surrogate objective:
\begin{equation}
\scriptsize
L_{\mathrm{AW\mbox{-}PPO}}
=
\mathbb{E}_{u,n}\!\left[
w_u^n
\min\!\Big(
r_u^n(\theta)\hat{A}_u^n,\,
\operatorname{clip}\!\big(r_u^n(\theta),1\!-\!\epsilon,1\!+\!\epsilon\big)\hat{A}_u^n
\Big)
\right]
\label{eq:awppo}
\end{equation}
The total training objective is given by
\begin{equation}
L_{\mathrm{total}}
=
L_{\mathrm{AW\text{-}PPO}}
+
c_1L_{\mathrm{value}}
-
c_2H[\pi_{\theta}],
\label{eq:total}
\end{equation}
where $L_{\mathrm{value}}$ is the value loss and $H[\pi_{\theta}]$ is the entropy regularization term. This formulation combines PPO's stable policy updates with AWAC-style advantage weighting, thereby improving learning efficiency and convergence under dynamic End--Edge--Cloud environments while ensuring stable cooperative learning across distributed agents.
\begin{figure*}[h]
\centering
\subfloat[]{\includegraphics[width=0.3\textwidth]{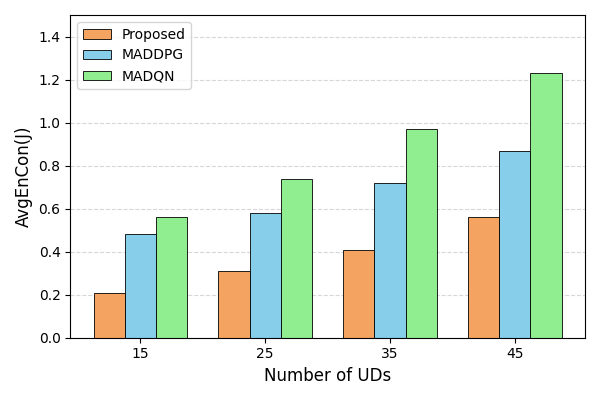}\label{fig:delay150}}
\hfil
\subfloat[]{\includegraphics[width=0.3\textwidth]{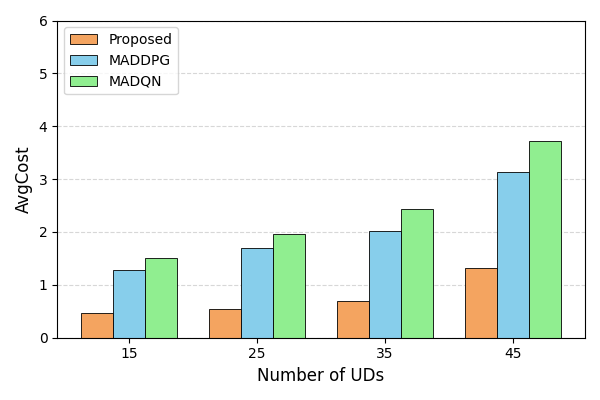}\label{fig:delay250}}
\hfil
\subfloat[]{\includegraphics[width=0.3\textwidth]{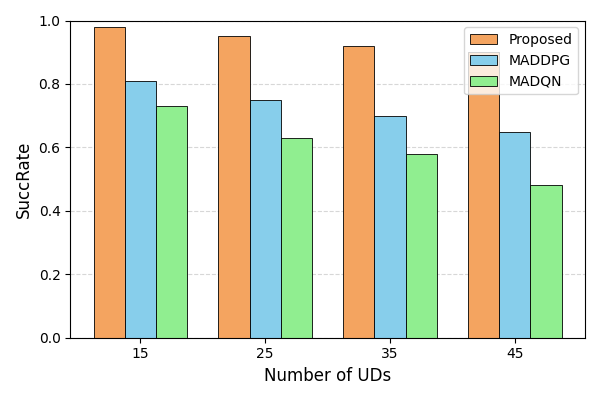}\label{fig:delay350}}
\caption{Performance comparison of the proposed framework and benchmark schemes under varying numbers of UDs in terms of (a) average energy consumption, (b) average cost, and (c) success rate.}\vspace{-3mm}
\label{fig5}
\end{figure*}
\section{Performance Evaluation}
\label{sec:results}
The proposed personalized privacy-aware End--Edge--Cloud continuum is deployed over a $0.5\times0.5~\text{km}^2$ area, where the number of UDs is varied from $10$ to $45$ to evaluate the scalability of the proposed framework. User trajectories are generated using a random-walk mobility model across multiple ESs. Each UD generates a computation-intensive and latency-sensitive task that can be executed locally or partially offloaded to ES and CS. For task $J_u^n$, the input data size $L_u^n$ is uniformly distributed within $\mathcal{U}(0.8,1.2)$ MB, while the required computational intensity is set to $C=1000$ cycles/bit. The system bandwidth is $B=1$ MHz, the receiver noise power is $\sigma^2=-100$ dBm \cite{wang2021computation}, and the uplink transmission power is $p^{tx}=0.1$ W. The local CPU frequency of UDs is selected from $f_u\in[0.6,1]$ GHz, whereas the ESs provide computational resources ranging from $f_s\in[4,8]$ GHz. Unless otherwise stated, the maximum tolerable delay is set to $D^{\max}\in[0.6,1]$ s. The CS is assumed to possess abundant computational resources compared with ESs.
The proposed framework is implemented in Python. For the AWAC-enhanced PPO-based MARL framework, the discount factor is set to $\gamma=0.99$, and the PPO clipping coefficient is set to $\epsilon=0.2$, following the implementation settings in \cite{chandler2025network}. The performance of the proposed scheme is evaluated under different network scales and privacy requirements in terms of average system cost, energy consumption, and task success rate.
Performance evaluation metrics include three types. The first is the \textbf{average system cost (AvgCost)} defined in \eqref{cost} of all successful episodes, which reflects the overall quality of service of task execution in the proposed End--Edge--Cloud framework. The second is the \textbf{average energy consumption (AvgEngCon)}, calculated from \eqref{eq:energy} and averaged over all completed tasks. The third is the \textbf{task success rate (SuccRate)}, calculated as the percentage of tasks completed within the maximum tolerable delay constraint, which can be expressed as
$\sum_{n=1}^{N}\mathbb{I}(t_u^{n}\leq D^{\max})/N$,
where $\mathbb{I}(\cdot)$ is the indicator function, $t_u^{n}$ denotes the realized latency of task $J_u^{n}$, and $D^{\max}$ is the maximum tolerable delay.

Baselines include two representative state-of-the-art MARL-based approaches widely adopted in MEC systems, namely \textbf{MADQN} \cite{li2018task} and \textbf{MADDPG} \cite{guo2022multi}, both following the CTDE paradigm. \textbf{MADQN} employs a discrete action space with quantized offloading levels $\{0,\,0.1,\,\ldots,\,1.0\}$, whereas \textbf{MADDPG} directly learns continuous offloading decisions.

Fig.~\ref{fig5} compares the performance of the proposed framework and the benchmark schemes under varying numbers of UDs in terms of average energy consumption, average system cost, and task success rate under personalized privacy requirements.

As shown in Fig.~\ref{fig5}(a), the AvgEngCon of all schemes gradually increases with the number of UDs, owing to the higher communication and computation demands under denser network conditions. However, the proposed framework exhibits only a moderate increase, whereas the energy consumption of \textit{MADDPG} and particularly \textit{MADQN} grows more rapidly. This indicates that, by incorporating the UCP model into the offloading decision, the proposed framework allocates privacy-sensitive tasks more judiciously and utilizes computing resources more efficiently, thereby reducing unnecessary transmission and processing overhead.

A similar trend is observed for the average system cost in Fig.~\ref{fig5}(b), where the AvgCost of all schemes rises as the network load intensifies. Nevertheless, the proposed framework consistently maintains the lowest AvgCost, and the gap relative to the benchmark schemes becomes more pronounced under heavy network loads. This behavior confirms that jointly optimizing the offloading ratio and target server while satisfying heterogeneous privacy constraints enables more efficient task offloading and resource allocation, effectively reducing the overall system cost without sacrificing privacy preservation.

In contrast, the task success rate in Fig.~\ref{fig5}(c) decreases for all schemes as the number of UDs grows, due to intensified competition for limited computing resources. Despite this, the proposed framework consistently sustains a SuccRate above 90\% across all evaluated network settings, while the success rates of \textit{MADDPG} and especially \textit{MADQN} decline much more noticeably. This demonstrates that, unlike the baselines that overlook user-specific privacy demands, the proposed framework provides privacy-aware task offloading that balances privacy protection with service performance, enabling a larger proportion of tasks to satisfy the required latency, energy, and privacy constraints.\vspace{-3mm}
\section{Conclusion}
\label{sec:conclusion}
This paper investigated personalized privacy-aware computation offloading in an End--Edge--Cloud continuum, where each UD minimizes the long-term system cost by jointly optimizing the target computing server and the partial offloading ratio under heterogeneous privacy requirements characterized through a UCP model based on task sensitivity, user privacy preference, and task urgency. To solve the resulting optimization problem, we developed a decentralized MARL framework based on an AWAC-enhanced PPO scheme, which combines PPO's stable policy optimization with AWAC-style advantage weighting to improve policy learning under dynamic network conditions. Extensive simulation results demonstrate that the proposed framework consistently achieves lower average system cost, lower energy consumption, and higher task success rate than the benchmark methods. These findings verify the effectiveness of integrating personalized privacy modeling with stable multi-agent policy learning for privacy-aware task offloading. In future work, we aim to develop a unified decision-making framework by jointly incorporating additional system information and scalable multi-agent coordination for large-scale deployments.\vspace{-2.5mm}
\section{ACKNOWLEDGMENT}
This work has received funding from the European Union under the EXACT-6G project (Grant Agreement No. 101120297) and the 6G-INTENSE project (Grant Agreement No. 101139266). \vspace{-3mm}
\bibliographystyle{IEEEtran}
\bibliography{bib.bib}
\end{document}